\documentclass[fleqn,usenatbib]{mnras}

\usepackage{mathptmx}
\usepackage{txfonts}

\usepackage[T1]{fontenc}

\usepackage{graphicx}

\DeclareRobustCommand{\VAN}[3]{#2}
\let\VANthebibliography\thebibliography
\def\thebibliography{\DeclareRobustCommand{\VAN}[3]{##3}\VANthebibliography}

\usepackage{graphicx}	

\title[Jet eclipses]{Eclipses of jets and discs of X-ray binaries as a powerful tool for understanding jet physics and binary parameters}

\author[T.J. Maccarone et al.]{Thomas J. Maccarone,$^{1}$\thanks{E-mail: thomas.maccarone@ttu.edu (TJM)}
Jakob van den Eijnden,$^{2}$
Thomas D. Russell$^{2}$ and
\newauthor
 Nathalie Degenaar$^{2}$
\\
$^{1}$Department of Physics \& Astronomy, Texas Tech University, Box 41051, Lubbock TX 79409-1051, USA\\
$^{2}$Anton Pannekoek Institute, University of Amsterdam, Amsterdam, The Netherlands\\
}

\date{Accepted XXX. Received YYY; in original form ZZZ}

\pubyear{2020}

\begin{document}
\label{firstpage}
\pagerange{\pageref{firstpage}--\pageref{lastpage}}
\maketitle

\begin{abstract}
We calculate the expected effects on the spectral energy distributions and light curves in X-ray binary jets from eclipses by the donor stars. Jets will be eclipsed for all inclination angles, with just the height along the jet where the eclipse takes place being set by the orbital parameters.  Typically, eclipses will lead to 5-10\% reductions in the jet emission over a range of a factor of few in wavelength with a periodic modulation.  In ideal systems with high inclination angles, relatively even mass ratios, and modest jet speeds, the eclipses may be deeper.  We discuss how eclipses can be used to measure binary system parameters, as well as the height of the bases of the jets.  We also discuss how, with data sets that will likely require future facilities, more detailed tests of models of jet physics could be made, by establishing deviations from the standard recipes for compact conical flat spectrum jets, and by determining the ingress and egress durations of the eclipses and measuring the transverse size of the jets. We provide representative calculations of expectations for different classes of systems, demonstrating that the most promising target for showing this effect in the radio band are the longer period "atoll"-class neutron star X-ray binaries, while in the optical and infrared bands, the best candidates are likely to be the most edge-on black hole X-ray binaries.  We also discuss the effects of the outer accretion disc eclipsing the inner jet.

\end{abstract}




\section{Introduction}

Historically, the best empirical data for studies of stars have come from the study of eclipsing binary stars (see e.g. \citealt{Southworth} for a recent review). In standard eclipsing binaries, the measurement of the kinematics using primarily spectroscopy, but also by timing the duration of the eclipses, can be combined with the measurements of the relative radii and temperatures of the two stars from the photometry, and can yield precise solutions for masses, temperatures and radii of the two stars as well as the distance to the binary.   The first two eclipsing binaries discovered, Algol and $\beta$ Lyrae \citep{Goodricke1783,Goodricke1785}, are now understood to be mass-transferring binaries, and the eclipse mapping studies provided the first clues that started studies of binary stellar evolution \citep{Kuiper1941,Crawford1955}.  Over time it has become clear that a substantial fraction of eclipsing binaries have undergone mass transfer at some point, since the binaries most likely to show both eclipses and mass transfer are the ones with the smallest ratio of orbital separation to the radius of the larger star.  Wider binaries can be used to understand standard stellar evolution, as they suffer neither from mass transfer nor mergers \citep{deMink2011} and the rare wide, eclipsing binaries still provide the opportunity to use eclipse mapping. 

Eclipse mapping also represents one of the core methods for probing the structure of accretion discs around compact objects (see \citealt{MaccaroneReview} for an overview of a range of techniques, and see \citealt{Baptista2001} for a review more focused on eclipse mapping). Here, both the vertical and radial structure of the accretion disk can be probed by looking at the brightness of the accretion disc component as a function of orbital phase.  

Even some of the earliest proportional counter-based X-ray spectroscopy of accreting black holes \citep{Tananbaum1972}, led to the understanding that the inner accretion discs around black holes could not be fully explained in all cases by the standard geometrically thin, optically thick accretion disc models \citep{ShakuraSunyaev}, thus requiring Comptonization-based models \citep[e.g.,][]{Thorne1975}.  With eclipse mapping studies, primarily done on cataclysmic variable stars, it became understood that even the outer parts of the accretion disks are not fully explained by these standard thermally emitting viscously-powered disc models \citep{Wood1992,Robinson1999}.  Eclipse mapping studies of neutron star X-ray binaries also revealed that the X-ray emission appears to originate from a spatially large region \citep{Church2004}, although these studies may, in fact, indicate that accretion disc winds which scatter X-rays produced closer to the neutron stars are ubiquitous in these systems \citep{Begelman1983}, which could, in turn provide evidence for disc winds even in spectral states dominated by hard X-ray emission.\footnote{If confirmed, this would suggest that the correlation between the spectral state of the accretion disc and its disc winds \citep{2009Natur.458..481N} may arise from ionization effects that prevent the disc winds from being easily detected in certain spectral states if absorption lines are used as the tracers of the disc winds.}

Relativistic jets from X-ray binaries are difficult to study definitively using spectroscopy, making eclipse mapping of these systems especially valuable if it can be made to work.  In most cases, jets have no spectral lines (although see \citealt{Margon1979,DiazTrigo2013} for a few exceptions).  Furthermore, jets are non-thermal, optically thin,  and lack spherical symmetry.  The jets appear to come in two types: steady compact jets, and rapidly evolving transient jets.  While the transient jets launched by X-ray binaries can sometimes be studied using imaging data, only in a small number of cases have the steady, compact jets be resolved \citep{Dhawan2000,Stirling2001,Russell2015}.  Given that the characteristic size scale of the jets increases linearly with wavelength, improved ability to make direct images of extended jet structures can be obtained only by increasing the longest baselines of the interferometer used.  Thus, since these sources are rarely resolved, even with continental-scale baselines, only the discovery of new closer systems or the use of space-based interferometry provide hope for improvement in our ability to image the compact jets from X-ray binaries.
\begin{table*}
	\centering
	\caption{The parameters that go into our model.  Note that we define $m_1$ to be the accretor mass and $m_2$ to be the donor mass regardless of which object is more massive.}
	\label{tab:jetparams}
	\begin{tabular}{ll} %
		\hline
		Variable & Description \\
		\hline
        $r$& radius of the donor star\\
        $a$& semi-major axis of the orbit\\
        $m_1$& mass of the accretor\\
        $m_2$&mass of the donor\\
        $\nu_{br}$&the break frequency at the base of the jet\\
        $h_{min}$& the jet base's distance from the center of the compact object \\
        $i$ &inclination angle of the orbit\\
        $\phi$ & orbital phase\\
        $\alpha$ & the spectral index of the jet above $\nu_{br}$\\
        $\theta$ & the opening angle of the jet\\
        $q$& binary mass ratio, $m_2/m_1$\\
         \hline
	\end{tabular}
\end{table*}

\begin{table*}
	\centering
	\caption{The key binary system parameters and how they are measured from the jet eclipse}
	\label{tab:binparams}
	\begin{tabular}{lll} %
		\hline
		Parameter & Description &  Measurable \\
		\hline
        $q$ & mass ratio & duration of eclipse\\
        $i$  & inclination angle & range of frequencies observed to be eclipsed\\
        $r_{circ}$ & the circularization radius of the accretion disc & upper limit from inner jet eclipse/correlation with $q$\\
        $r_{out}$ & the actual outer radius of the accretion disc & innet jet eclipse and correlation with $q$\\
        \hline
	\end{tabular}

\end{table*}
The compact jets are reasonably well-explained by models in which at each height in the jet, a moderately steep spectrum synchrotron component is emitted, with synchrotron self-absorption setting in at a wavelength that increases linearly with distance moving away from the compact object powering the jet \citep{BK1979}.  This dependence then means that the {size scale of the jet in units of the angular resolution of an interferometer with fixed maximum baseline length} is independent of the wavelength of light used to study it.  Since very long baseline interferometry (VLBI) arrays already span nearly the whole radius of the Earth, only with space-based VLBI is it plausible to resolve the jets which cannot currently be resolved.  The jets often show mildly inverted {($\alpha = 0.0-0.2$)} radio spectra \citep[e.g.,][]{Fender2001,Corbel2013}.  In AGN, inverted spectra have been explained by free-free absorption from local gas \citep{Walker2000}, but this issue has not been well-studied in the X-ray binaries.  It thus may be possible that the inverted spectra result from the jet geometry, the bulk velocity along the jet (making its Doppler factor closer to unity; e.g., \citealt{BK1979,Falcke1995}) or loses power as it propagates due to , e.g., shocks and turbulence \citep{Kaiser2006,Jamil2010}.  Deviations from the model predictions in this paper could be used to test scenarios for the inverted optically-thick part of the spectra. In this paper, we simply note that the issue exists, but could consider a range of possibilities for the inverted spectra in future work.

Given the lack of diagnostic power from imaging and spectra, alternative, timing-based, approaches to resolving the structures of the jets are thus necessary to move forward with our understanding of relativistic jets in X-ray binaries.  Studies presenting time lags of jet emission behind the X-ray emission that comes from the accretion disc have already started to produce constraints on jet speeds and jet acceleration \citep{Casella2010,Gandhi2017,Tetarenko2019}.  Past work has shown orbital modulation of the radio emission in Cygnus X-1 and Cygnus X-3 \citep{Brocksopp2002,Szostek2007,Zdziarski2018}, but in these high mass X-ray binary systems, the eclipsing is not due to the donor stars themselves, but rather the winds from the donor stars.

Observing eclipses of relativistic jets from X-ray binaries by their donor stars is important for a few reasons.  First, it gives the potential to understand the structure of the relativistic jet.  The transverse sizes of X-ray binary jets are far beyond the capability of any proposed imaging facilities.  Even with the eclipse mapping technique, it is likely to be quite challenging to measure the transverse sizes, as in most cases, the jet's transverse size scale will be significantly smaller than the donor star's radius, meaning that the ingress and egress of the eclipse will be much shorter than the eclipse.  Still, the advantage of periodic phenomena based is that, via phase-folding and/or Fourier methods, signal-to-noise can be built up gradually; unlike imaging, eclipse mapping gives a process by which geometric estimates of the jet width can be made.

Second, it gives the opportunity to understand the binary system parameters.  First, for many X-ray binaries, the orbital periods are not yet known.  Orbital periods can often be measured using the donor stars' ellipsoidal modulations in quiescence \citep{Avni1975}.  For systems which are face-on, however, these ellipsoidal modulations can be very weak.  Furthermore, for systems which are persistently bright, the dilution of the ellipsoidal modulations by the accretion light can make the ellipsoidal modulations undetectable.  In all cases, the measurement of the inclination angles from ellipsoidal modulations requires careful estimation of the level of emission from the accretion disc and jet; this can be done, but it requires very large photometric data sets with good calibration \citep[e.g.,][]{Cantrell2010}.  On the other hand, orbital periods and inclination angles can be estimated in outburst using jet eclipses, and the quality of measurements should be relatively insensitive to the inclination angle.

Furthermore, there is substantial direct evidence that some X-ray binaries show ``misalignments'' between the angular momentum vectors of the orbit and of the accretor.  Evidence for these misalignments comes in the ubiquitous, but strongly model-dependent (and hence inconclusive) form of quasi-periodic oscillations (QPOs) that can be well-explained by the Lense-Thirring precession \citep{StellaVietri,Ingram2012,IngramReview}.  Even if the Lense-Thirring model is found to explain the "Type C" QPOs, the angular scale of misalignment is not something which can be cleanly determined from the amplitude of oscillations, so additional means for finding the actual angular size of the misalignment would be needed.  This may come from X-ray polarization \citep{Ingram2015}, but this has not yet been clearly established.  Two-sided jet proper motions, along with accurate distances, can give good estimates of the inclination angles of jets (see \citealt{Fender2003} for a discussion of this technique, including its limitations), although in V404~Cyg in its recent likely super-Eddington outburst, it showed changing inclination angles \citep{2019Natur.569..374M}.  

Some direct evidence for misalignments does, in fact, exist in a few systems, where the jet inclination angles can be estimated by two-sided proper motions, or constrained by highly superluminal motions, and are inconsistent with the orbital inclination angles estimate by ellipsoidal modulations (see e.g. \citealt{2002MNRAS.336.1371M, Beer2002,Hjellming1995}).  In one case, Cygnus X-3, additional evidence for misalignment comes from a phase offset between the jet {{flux}'s modulation by the donor star and the orbital phasing provides additional evidence for misalignment \citep{Dubus2010}.

Crucially, if combined with a good model for the jet's structure, the phasing of the jet {flux}'s modulation due to eclipsing by the donor star can reveal an offset in position angle of the jet relative to the orbital plane.  Given that the position angle of the polarization of the optically thick component of the jet should be the same as the position angle of the jet's direction of elongation \citep[e.g.][]{Zdziarski2014}, this potentially opens up the possibility of making an estimate of the position angle of the binary, even in the absence of any direct imaging information.  

A key application of being able to make such measurements is to be able to determine the relative directions of the proper motions and the orbital plane and jet axis directions for X-ray binaries which have undergone substantial natal kicks.  The kicks can come from the \citet{Blaauw1961} mechanism -- recoil due to the sudden loss of mass from a moving binary progenitor -- or from some additional natal kick \citep{Gunn1970}, the origins of which remain controversial.  Crucially, the "Blauuw" kicks should be entirely in the binary orbital plane.  Establishing that the jet is aligned can then yield the binary's orbital plane direction, rather than just the jet's, and this can be done using polarization information, which remains sensitive even for binaries which cannot be resolved directly.  Substantial kicks not consistent with this plane would then provide clear evidence for re-alignment of the accretor's spin with the orbital plane on a timescale faster than the evolutionary timescale for the binary.

In this paper, we lay out calculations for the expected properties of eclipsed jets.  We start by describing the jet model we use for the calculations (which is relatively standard, following \citealt{BK1979}).  We then present analytic expressions to best determine at which wavelengths the effects of jet eclipses are most likely to be measurable for which sources.  We present some representative examples of what the jet eclipses might look like under our best-guess assumptions for system parameters in order to illustrate in which frequency bands different classes of sources are most likely to eclipse and how deep typical eclipses should be.  We close the paper by discussing specific observations that can be made with current facilities that may prove enlightening about the nature of jets and system parameters for X-ray binaries as well as which future facilities would be most helpful for improving our understanding of these systems.  In this paper, we will adopt several simplifying assumptions, but we attempt to outline the likely qualitative effects of relaxing these assumptions and prospects for future refinement of the model.

\section{The jet model}

To estimate the effects of jet eclipses, we first construct a model of a compact conical jet, drawing from \citet{BK1979}.  A listing of the parameters which go into the jet model is given in Table \ref{tab:jetparams}, and a listing of the parameters of the binary orbit which can be estimated using jet eclipses, but which are not direct input parameters of the model, is given in Table \ref{tab:binparams}.  

We start from a jet model in which at each height within the jet, the spectrum is a broken power law, with a flux density, $F_\nu$, that is proportional to $\nu^{-0.7}$ above some break frequency and to $\nu^{+2.5}$ below that break frequency {with a constant brightness temperature at the peak and a conical structure}.  The former corresponds to typical jet spectra at high frequencies, and is well-explained as the synchrotron spectrum due to electrons accelerated by Fermi acceleration \citep{RandL}.  The latter is the characteristic spectral shape due to synchrotron self-absorption \citep{RandL}.  {The spatially differential flux density (i.e., $\frac{dF}{dh d\nu}$), where $F$ is the total flux, $h$ is the height along the jet, and $\nu$ is the frequency), is assumed to peak at a level inversely  proportional to the height along the jet and to peak at a wavelength that is linearly proportional to the height.}  It is assumed that there is no jet emission below the base of the jet.  Alternative models to explain flat spectra would require correlations of jet spectral shape and jet power with height along the jet, and hence it is unlikely that such models apply to the real sources, but if they do, they will produce clear deviations from the model predictions for jet eclipses presented in this paper.

{We make the initial calculations without special relativistic corrections, and apply those corrections as flux enhancements at the end of the calculation to account for approaching and receeding components' different doppler boosting.  This is discussed in more detail below. We assume that the height at which the jet becomes optically thick depends only on distance along the jet and not on the inclination angle, but we do properly project the jet as a function of inclination angle.  We do not account for the cutoff of the jet's spectrum due to synchrotron cooling, but for the purposes of this paper the effects will be negligible.}

This scenario results in a $F_\nu \propto \nu^0$ spectrum with a break to a $\nu^{-0.7}$ spectrum above the break frequency that exists at the bottom of the jet, where the particles are first accelerated.  For many systems, the low/hard state jets show slightly inverted spectra, with $F_\nu \propto \nu^{0.0-0.4}$, and, in rare cases, even more inverted. The model can be adapted to match this phenomenology by making the normalization of the peak of the spectrum increase slightly toward the base of the jet.  We do not consider such an effect in this paper, but it is straightforward to adapt the model in that manner in future work.      

To calculate a jet spectrum, we break the jet into segments, $\Delta{\rm{log}~ h}$ in size.  At each height, we test whether the donor star is between the viewer and the jet.  If so, this bin is excluded from the summation.  We additionally consider whether the outer accretion disc may eclipse the jet. We present some calculations where we consider this effect, to show the full effects of eclipsing on the jet spectra.  We also present some calculations where we show only the effects of the star eclipsing the jet, because the only the star's eclipses will lead to orbital modulation.

We produce summations for both the approaching and receding (counterjets) components of the jets.  For the sample calculations below, we focus on neutron star jets.  For these systems, the relativistic effects are likely to be quite mild.  For example, if neutron stars typically launch jets at their escape speeds (which are $\sim 0.3c$ for typical equations of state -- e.g. \citealt{Steiner2010}), then for a 45 degree inclination angle, the relativistic effects will give a factor of $\sim2$ or less enhancement of the approaching jet relative to the counterjet.  Direct measurements of expansion speeds in Sco~X-1 are $\sim0.4\pm0.1c$ in line with this value \citep{Fomalont2001}.  In a few cases, some evidence for {\it energy transport} has been seen that may occur at much higher speeds \citep{Fomalont2001, Fender2004}, based on associations of time-delayed brightenings in the jet with brightenings in the core. There is no clear evidence that the gas itself is moving at such high speeds, and instead, these could just be due to Poynting flux transport of energy, or even mis-associations between core events and events further out in the jet.  It is thus quite reasonable to expect that the jets from neutron stars will generically be only mildly relativistic, so that the jet/counterjet flux ratio will be relatively small.

For black hole jets, where the speeds are often seen to reach {$\sim0.8-9c$} or more (e.g., \citealt{Mirabel1994,Tingay1995,Corbel2002}, see \citealt{Fender2006} for review, \citealt{Saikia2019} for a more recent compilation), the approaching jets can be expected to be tens or more times brighter than the counterjets over a wide range of inclination angles. While the counterjet will always be eclipsed at some height, in most cases, the approaching jet will be completely uneclipsed.  Eclipsing of the approaching jet will take place only for systems which are quite edge-on, and which should then also show eclipses of the X-ray emission from the accretion disc, something no dynamically confirmed Galactic black hole X-ray binary does.  Therefore, the effects we discuss in this paper are likely to be quite subtle in black hole X-ray binary systems unless the system is either: (1) nearly edge-on, (2) has a sufficiently strongly misaligned jet that the approaching jet is eclipsed, or, alternatively, (3) the jet is significantly slower than what has been estimated in past work.  

The possibility of jets moving at half the speed of light or less from some subset of black hole accretors cannot be discounted.  Considerable evidence exists for slower jet speeds at low power than at high power.  In most X-ray binaries, bright radio flares are observed when sources transition from their hard states to their soft states, likely because the pre-existing hard state jet provides a work surface for the faster-moving transient jet, which dissipates its power effectively against a large mass of material \citep[e.g.,][]{Vadawale2003,Fender2004}. Such an argument is bolstered by the fact Cygnus X-3 is the only source which shows brighter jets at the transition from the soft state to the hard state than during the hard-to-soft transition. For Cygnus X-3, this is thought to arise from the work surface for the jet being the stellar wind from the companion. During the hard states, the jet drills out a cavity, which is filled back in during the soft state, meaning that when the hard state jet re-ignites, the work surface is present close to the compact object \citep{Koljonen2018}.  

Finally, the variation in the radio/X-ray correlation across different systems suggests that the jet speeds may vary substantially, with $\Gamma$ {perhaps as much as} $\approx 3-4$ for bright hard states, and the jets being only very mildly relativistic for lower luminosities \citep[e.g.,][]{Russell2015}.  {Other evidence for jets at about this speed comes from radio timing of Cygnus~X-1 \citep{Tetarenko2019}.  Conversely, the relative lack of scatter in most of the radio/X-ray relation has been argued to indicate typical Lorentz factors less than 1.7 \citep{2003MNRAS.344...60G,Soleri2011}, although it has also been shown that as long as the range of Lorentz factors is small, higher speeds can be tolerated \citep{HeinzMerloni2004}.}. Thus, while direct speed measurements of the faint jets do not yet exist, considerable evidence points toward the faint hard-state jets being less relativistic.  {The evidence for the high jet speeds in bright hard states is still presently tentative and based on a relatively small number of sources, so it will be worthwhile to test these measurements on some black holes, as well, but on the whole, we are more optimistic about this work being successfully applied to neutron star systems.} Additionally, in some cases, jets may slow down from than their original ejection speed as they entrain material or interact with their surrounding environment \citep[e.g.,][]{Corbel2002,Kaaret2003,Russell2019,Bright2020}, for example, strong emission lines coming from the jet in SS~433 revealed the jet entraining material \citep{Margon1979}.  In such a case, the jet model discussed here is likely to break down in several other ways, and more {\it ad hoc} modelling approaches will be needed.

After setting up our initial model, we run the code for a range of orbital phases, allowing us to sample the eclipse pattern.  We will present both spectra at different orbital phases, and light curves at different frequencies.

\section{The input parameters: general considerations}

The key input parameters for the jet eclipse model are the orbital parameters for the binary system and the jet system parameters.  The binary system parameters will be discussed on a case-by-case basis as we present predictions for individual sources.  We use a mixture of theoretical and observational approaches in this paper to determine the appropriate parameters for calculations.  We attempt to discuss where our assumptions may break down and what the consequences of this may be.

\subsection{Spectral index}
We regard the spectral index of the emission above the break to be well-estimated by observations of active galactic nuclei \citep{Readhead1979,Eckart1986,Hovatta2014}, and work under the assumption that the same acceleration physics is present in the X-ray binaries.  Furthermore, the model predictions are only weakly sensitive to the spectral index of the jet.  We thus take $F_\nu \propto \nu^{\alpha}$, and set $\alpha$ to $-0.7$, and assume that the spectrum of each small region of the jet is also a $\alpha=-0.7$ power law above the frequency where it becomes optically thin.  

\subsection{Break frequency}
For a given jet, the free parameters are the inclination angle, the height of the base of the jet, the offset in direction relative to that of the orbit, and the overall flux normalization.  There are theoretical predictions relating the frequency of the jet spectral break, $\nu_{br}$ to the kinetic power in the jet, $P_{jet}$.  \citet{HeinzSunyaev2003} show that $\nu_{br} \propto P_{jet}^{2/3}$.  For most cases where there are compact jets produced by black holes, the X-ray luminosity, $L_X$, is expected to scale as $\dot{m}^{2}$, due to advection-dominated accretion flows \citep{1995ApJ...452..710N}, while the jet kinetic power remains linearly proportional to the accretion rate \citep{Meier2001}.  For radiatively efficient accretion onto neutron stars $L_X$ should be proportional to $\dot{m}$.  Thus, for accreting black holes in their low luminosity states, we expect that $\nu_{br} \propto L_X^{1/3}$, and for accreting neutron stars, $\nu_{br} \propto L_X^{2/3}$ for steady hard states, where it is most practical to make observations of jet eclipses.

The brightest black hole X-ray binaries and neutron stars in hard states with steady jets tend to be at 2\% of the Eddington luminosity {in the more slowly evolving decaying hard states} \citep{Maccarone2003,Armin2019}.  {In the more short-lived rising phase of outbursts, hard states can be seen up to $\sim$ 20--30\% of the Eddington luminosity} \citep{Miyamoto1995,2003MNRAS.338..189M}; {these very bright hard states would potentially make excellent targets for jet eclipse studies, but are operationally more challenging for triggered observations}.  Quiescent stellar mass black hole X-ray binaries tend to be at luminosities of $10^{-9}-10^{-6}$ of the Eddington luminosity \citep{Plotkin2013}, meaning that the break frequencies should be expected to move by a factor of 20--200.  

Empirically, such a large range in estimated break frequency is seen, but the correlations between break frequency and luminosity are highly scattered \citep{Russell2013,Russell2014}.  It is worth noting that the breaks are often not directly detected because often, and especially for the faintest sources, the jet is fainter than the donor star or the thermal emission from the outer accretion disk at the break frequency. Additionally, where (likely) robust repeated measurements from the same source during their outburst phases have been made, the break frequencies have been seen to move substantially over a small range in X-ray luminosity (\citealt{Russell2014}, but see also \citealt{Corbel2013} and \citealt{vanderHorst2013}). However, such an evolution was occurring around and over spectral state transitions during outburst, where the structure of the jet (and accretion flow) may be evolving quickly.  

Using spectral decomposition to estimate the jet spectrum in the optical/infrared regime where often all three components contribute (as well as, perhaps, a synchrotron emission from the inner hot flow -- \citealt{Veledina2013,Kosenkov2020}, or the pre-shock region of the jet -- \citealt{Markoff2001}), can be hampered by uncertainties in reddending and irradiation of the outer disc.  On the other hand, the \citet{HeinzSunyaev2003} relations predict the fundamental plane of black hole activity \citep{Merloni2003}, which relates the X-ray and radio powers with a black hole mass scaling.  {Even the standard radio/X-ray relation for stellar mass black holes \citep{2003MNRAS.344...60G} requires } there be either a fine-tuning of parameters like the jet opening angles to allow the break frequencies to vary in a manner far different from the \citet{HeinzSunyaev2003} predictions while the fundamental plane quantities follow its predictions, or the severe observational challenges in estimating the break frequencies make the theoretical prediction of the breaks more reliable than the observational attempts to estimate them.  

For the purposes of this paper, we work on the basis of the idea that the theoretical predictions are more reliable, but we note that the jet eclipse measurements, both by the donor and by the outer disk, can help break this degeneracy in a manner quite independent of other approaches in the literature.  A core qualitative issue is that inner hot flows \citep{Veledina2013} should be eclipsed by the donor stars if and only if the X-ray emission is eclipsed, while the inner regions of the jet can be eclipsed in some cases when the inner hot flow is not.  Additionally, eclipses of the inner hot flow by the outer parts of the disk should happen only for extremely edge-on systems with flared outer discs, while eclipses of the inner counterjet by the outer disk should occur for a broad range of inclination angles.  If sufficiently good data can be obtained to establish whether such eclipses are taking place, then this may provide a key for understanding which parts of the infrared emission come from the inner hot flow versus the jet.

\subsection{Height of the base of jet}
The heights of the jets above the accretor are also studied in the literature, with some clear progress having been made, but little that is definitive.  Observational evidence can come in the form of imaged jet size scales at particular frequencies, and timing-based estimates of the jet size scales at particular frequencies.  

The time delays to the optical and infrared emission regions in black hole X-ray binaries are typically 0.1-0.2 seconds \citep{Casella2010,Gandhi2017}, and the Lorentz factors of the jet in this regions is $\Gamma>2$, at least for GX~339-4 \citep{Casella2010}.  The jet bases are then likely to be $\approx3\times10^9$cm in size for stellar mass black holes, since the infrared and optical bands where these measurements are made are generally quite near $\nu_{br}$.  In making these estimates, \citet{Casella2010} also took into account the imaging data for Cygnus X-1.  

From radio timing data of Cygnus~X-1, \citet{Tetarenko2019} were able to make estimates of the jet speed and the relation between wavelength and timescale.  There, in the context of a model with no acceleration of the jet, it is found that $\Gamma = 2.59^{+0.79}_{-0.61}$, with the jet size scale proportional to $\nu^{-0.4}$.  This can, alternatively be explained if the jet is being accelerated through the emission region.  Furthermore, the model of \citet{Tetarenko2019} does not fully consider the effects of opacity from the stellar wind of the donor star, which also varies with height, on the emission properties \citep{Zdziarski2012Cygx1}.  Thus, until the system is re-modelled, we take these results as tentative, and focus on the default theoretical model that reliably predicts the jet's flat spectrum nature and the fundamental plane relation.

For the neutron star jets, the most natural interpretation will be that the characteristic spatial scale will be a factor of about 5 smaller, given the mass scaling linearly to the $1.4 M_\odot$ neutron stars, so a value of $\sim6\times10^8$cm can be assumed.  Given the slower speeds of the neutron star jets, this may, then predict only slightly shorter time lags for the jet base.  Unfortunately, no estimates of these time lags presently exist, and no source shows resolved steady jets\footnote{Two sources, Sco~X-1 \citep{Fomalont2001} and Cir~X-1 \citep{Coriat2019} show moving jets in, or shortly after, flaring states}.  The break frequencies on the other hand, have been directly estimated at least for the ultra-compact X-ray binary 4U~0614+091 \citep{Migliari2006,Migliari2010}, Aql~X-1 \citep{DiazTrigo2018}  and 4U~1728--34 \citep{DiazTrigo2017}.  For systems where the break frequency is known, but the base height of the jet is not known, eclipse mapping has the potential to provide an estimate of the base height.

\section{Analytic approximations, considerations for testing the model and choice of sources}

In order to develop some intuition for the expected results for different parameter values, and to determine which facets of the problem can be studied robustly, we present some analytic estimates of the properties of jet eclipses.  The jet model relies on the parameters in Table\ref{tab:jetparams}, which we present to help the paper be easy to understand.  In Figure \ref{fig:schematic} and Figure \ref{fig:schematic2}, we present schematic views to help illustrate the key geometric issues in jet eclipsing.

\subsection{Eclipsed region missing flux}
{
We can provide analytic expressions for the jet eclipse, although these are rather unwieldy and may become difficult to adapt to cases where the jet's opening angle or speed might change with height.  The jet's spectrum will be produced by summing two components: the optically thick and optically thin components.  The total flux density of an uneclipsed jet will be given by:}
\begin{equation}
    F_\nu =  \left(\frac{F_0}{2.5}\right)\left[1-\left(\frac{\nu}{\nu_{br,min}}\right)^{2.5}\right]+\left(\frac{F_0}{0.7}\right)\left[1-\left(\frac{h_{max}}{h_{br,\nu}}\right)^{-0.7}\right],
\end{equation}
{where $F_0$ is an overall normalization, approximately 1/1.8 of the flux density of the flat spectrum component; $\nu_{br,min}$ is the break frequency at the base of the jet, $h_{max}$ is the maximum height in the jet, and $h_{br,\nu}$ is the height at which the break frequency between optically thick and thin is $\nu$.}

{To determine the flux density associated with an eclipse, one then needs to calculate the same integral used to produce the equation above, with limits associated with the star's size, or the range of heights eclipsed by the disk. } {This produces the result:}
\begin{equation}
    F_\nu =  \left(\frac{F_0}{2.5}\right)\left(\frac{\nu}{\nu_
{br,min}}\right)^{2.5}\left[\left(\frac{r_{low}}{h_{min}}-1\right)^{2.5}\right]+\left(\frac{F_0}{0.7}\right)\left[1-\left(\frac{r_{high}}{h_{br,\nu}}\right)^{-0.7}\right],
\end{equation}

where $r_{low}$ and $r_{high}$ {are the lowest and highest locations in the eclipsed portion of the jet.  We do not use these expressions for the implementation in our code, but rather integrate numerically.}

\begin{figure*}
    \centering
    \includegraphics[width=6 in]{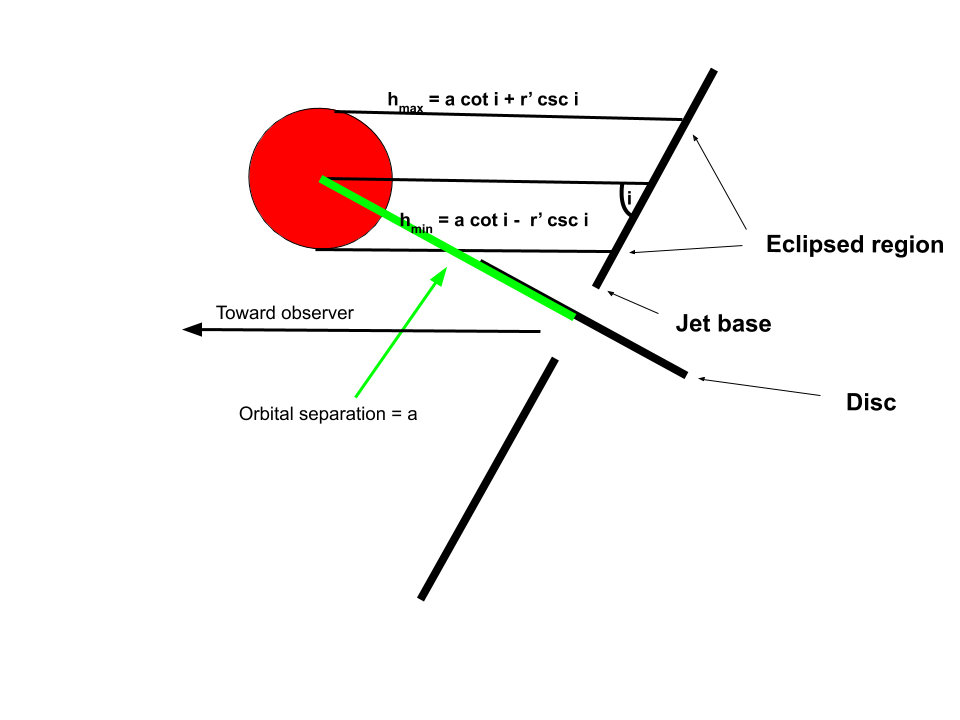}
    \caption{A schematic diagram to help illustrate the system geometry for a jet eclipse in a binary with inclination angle $i$.  The lines intersecting the jet and star represent rays pointed toward the observer.  The region from $h_{min}$ to $h_{max}$ is the eclipse region.  The $h_{base}$ region gives the height of the base of the jet where substantial emission starts.}
    \label{fig:schematic}
\end{figure*}

\begin{figure}
    \centering
    \includegraphics[width=3.5 in]{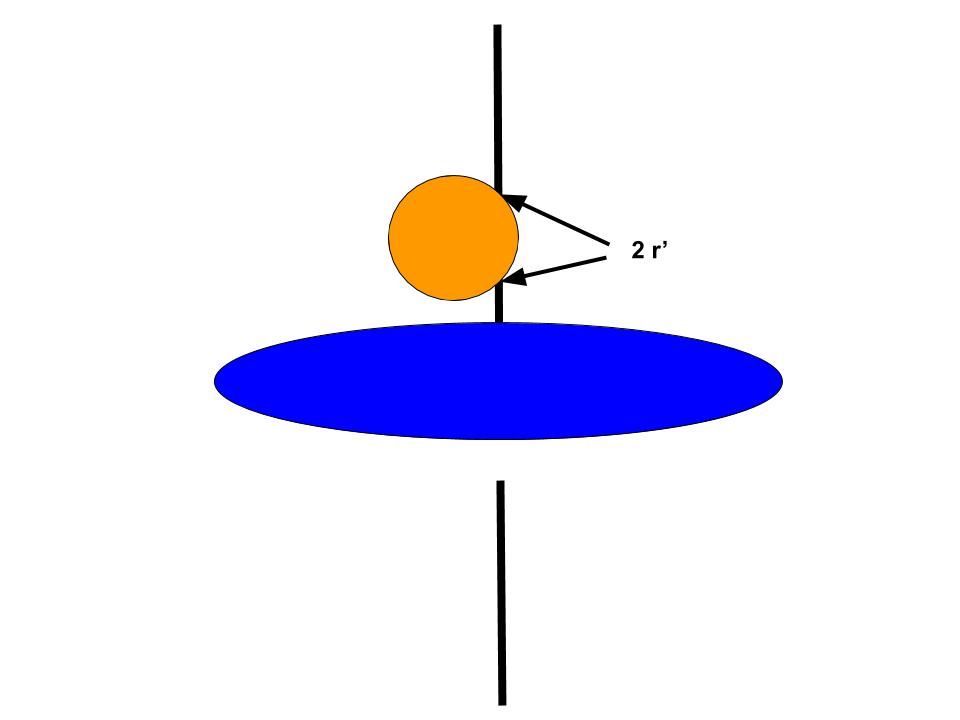}
    \caption{A schematic diagram drawn from the observer's perspective to illustrate the system geometry for a jet eclipse in a binary with inclination angle $i$.  In this figure, the accretion disk is eclipsing the inner counterjet, while the star is eclipsing a region further out in the counterjet.  The $r'$ represents the chord length for the part of the projected area of the star that blocks the jet.}
    \label{fig:schematic2}
\end{figure}

\subsection{Heights and range of wavelengths where the jet is eclipsed}
We can then derive some analytic expressions to determine where in the jet eclipses will take place and over which range of orbital phases.  The jet will be eclipsed where the height along the jet is in the range: 
\begin{equation}
    a~{\rm cot}~i-r'~{\rm csc}~i < h < a~{\rm cot}~i+r'~{\rm csc}~i.
\end{equation}  

The value $r'$ is equal to the half-length of the chord {that represents the eclipsed portion of the jet}.
.  This is illustrated in figure \ref{fig:schematic2}

This, then, gives an approach to estimating the inclination angles in the jet model.  Under the assumption that the jet is purely consistent with the assumptions that go into the flat spectrum model, then:
\begin{equation}
    \frac{\Delta \lambda_j}{\lambda_j} = \left(\frac{2r}{a}\right) {\rm sec}~i,.
\end{equation}
where $\lambda_j$ is the wavelength of emission from the jet.\footnote{Here, we express the equation in wavelength, rather than frequency units because the wavelength-based formalism is significantly simpler.}
The inclination angle can thus be determined from the range of frequencies which show some orbital modulation due to the eclipsing process.  The value of $r/a$ will come from the eclipse duration, so the inclination angle can be solved for directly.  In practice because we do not have all the light at any given wavelength coming from a single height in the jet, there will be some smearing of the effects even for the flat spectrum model, and with non-flat spectrum jets, or jets that show flat spectra due to coincidences in how parameters vary, things become even more complicated; still, there is a straightforward connection between the spectrum of the missing flux at mid-eclipse and the jet and the inclination angle, with large values of $\Delta \lambda_j$ corresponding to large inclination angles.

\subsection{Eclipse durations and constraints on the mass ratio of the binary}
 Understanding the mass ratios of binaries is necessary both for converting the binary mass function into a precise mass estimate and for understanding the evolution of the donor stars to determine the extent to which their structures deviate from those of main sequence stars filling Roche lobes of the same orbital period. At the present time, the primary techniques for obtaining the mass ratios of the binaries are measuring the rotational broadening of the donor star; \citealt{WadeHorne1988}, and measuring the separation of the double peaks of emission lines \citealt{Casares2016} relative to the full-width half-maximum of the lines.   The rotational broadening method requires high resolution spectra of the donor's absorption lines.  The latter requires a large number of spectra spread over a range of orbital phases to average out deviations from azimuthal symmetry in the disc.  For an aligned jet, the duration of the eclipse at the central frequency of the eclipsed part of the jet will depend only on the mass ratio of the binary (as shown below), so jet eclipses represent an alternative method. 

The jet eclipses will allow both controls on the systematics of the existing methods and sensitivity to different sets of systems.   The estimation of rotational broadening requires comparison with template spectra.  This leads to a potential source of systematic error for X-ray binaries, where the structures of the donor stars are often affected by ratios of mass transfer timescales $M/\dot{M}$, where $M$ is the donor mass and $\dot{M}$ is its mass transfer rate, that are shorter than their thermal timescales.  Under these circumstances, the stars are bloated due to being unable to relax to their equilibrium structure models quickly enough \citep{Rappaport1982}. The template stars, on the other hand, are generally single stars in equilibrium.  The double-peaked emission lines are susceptible to effects of both stochastic variability in the accretion disc and orbital modulation of the line profile, and hence require large numbers of measurements.  Furthermore, the relation between peak separation and mass ratio is established only in quiescence, and often, in outburst, the emission lines are single-peaked, probably due to scattering in strong disc winds \citep{Eracleous1994}.  The single peaked lines have been established to exist even in some systems where the accretion disc is in a hard state \citep{Wu2001}, arguing in a model dependent manner than the winds and jets can exist simultaneously, even outside the brightest states, in contrast to some of the conventional wisdom about these phenomena; regardless, in a model independent manner, it is clear that this technique cannot be used in bright accretion states, regardless of whether they are bright hard states or soft states.  Bowen fluorescence lines may also potentially be useful for tracing the donor stars' radial velocities and rotational broadening\citep{Steeghs2002,Hynes2003}, but at the present time, masses derived from these measurements have poorly understood systematics -- see, for example, the difference between the mass function for the black hole X-ray binary GX~339$-$4 inferred from Bowen fluorescence \citep{Hynes2003} relative to that from absorption lines \citep{Heida2017}. 

The use of the jet eclipses to estimate mass ratios is thus of greatest value for persistently bright sources, where other techniques are unlikely to work effectively (and this is also the case for low-power sources, where the jet may not be bright enough to allow straightforward observation of the eclipses anyway).  The jet will be eclipsed when:
\begin{equation}
    |{\rm sin}~(2\pi\phi)| < \left(\frac{r}{a}\right) 
\end{equation}
where $\phi$ is the orbital phase measured in the conventional manner, between 0 and 1.  For ultracompact X-ray binaries, with $q\sim0.03$, this will give eclipses for $-0.023<\phi<0.023$, while for systems with $q=0.5$, this will give eclipses in the range from -0.05 to 0.05.  This analysis treats the jet as a one-dimensional object and the star as a on object with a rigid surface.  Scale heights of stellar atmospheres are typically $\sim10^{-4}$ of the stellar radius, so this choice is unlikely to be a serious problem; only in cases where there is a strong stellar wind will the smearing due to the diffuse nature of the absorber be problematic.  

\subsubsection{Effects of the transverse extent of the jet}
The jet width is potentially a more serious problem, especially in systems with very low inclination angles, so that the eclipse takes place at relatively large heights in the jet relative to the donor star's radius.  Of course, if the jet width affects the duration of the eclipse, then, with high enough quality data showing a gradual ingress timescale, the jet's transverse size can be measured.  Very few jets from X-ray binaries have been resolved spatially in the transverse direction, and the measurements and upper limits indicate that the typical values are, at most, a few degrees \citep[see, e.g.,][]{MillerJones2006}.  The fractional increase in the eclipse duration can be determined by:
\begin{equation}
    \frac{\Delta r}{a}= \frac{{\rm tan}~(\theta/2)}{{\rm tan}~i},
\end{equation}
where $\theta$ is the jet opening angle.

The jet will eclipse the star for the same range of phases near $\phi=0.5$, but the brightness temperature of the star will be dramatically lower than the brightness temperature of the jet, so the eclipsing will have negligible effect.  Furthermore, in most cases, the jets will be optically thin in the optical and infrared bands where the star emits \citep[e.g.,][]{Russell2013}.  Alternatively, the high energy particles in the jet might potentially Compton upscatter photons from the star, but only in the cases of very large beaming factors or finely tuned misaligned jets is the flux from the donor star likely to be larger than the flux from the inner accretion disc in the inner region where the largest amount of power is dissipated, and even then, the jet's own synchrotron emission will likely still dominate the seed photons unless there is a well-tuned misalignment that boosts the star's photons dramatically (and in this case, the jet would likely not pass in front of the star at phase 0.5).

The caveats for use of the duration of jet eclipses to estimate the mass ratio in the binary are: (1) potential misalignment of the jet (2) measurement uncertainty (3) wavelength smearing and (4) effects due to the transverse size of the jet.  For jets which are misaligned, but straight, the offset in phase angle can be used to estimate the degree of misalignment, but a detailed model for the structure of the jet would be needed to extract a robust estimate of the inclination angle.


\section{Eclipsing of the inner counterjet by the outer accretion disc}
The inner portion of the counterjet may also be continuously blocked by the outer part of the accretion disc, with orbital modulation only if there is some deviation from azimuthal symmetry of the disc.  We will refer to this as the disc eclipsing the jet.  For aligned jets (i.e. jets perpendicular to the binary orbital plane), this will happen preferentially for systems that are closer to {\it face-on}.  This gives a potentially very powerful tool for estimating the jet inclination angles, although the magnitude of this effect will be relatively weak in face-on systems due to Doppler boosting unless the jet speeds are quite low.  

Similarly, the outer accretion disc can sometimes be eclipsed by the donor star even when the inner disk is not eclipsed, which can lead to grazing eclipses in the optical without X-ray eclipses.  The eclipsing of the outer disc {\it would} be periodically modulated.  This is not relevant to the calculations presented in this paper, but for more edge-on systems, it does present a potentially important potential additional source of periodic modulation, and consideration of this possibility is required if there is a periodic modulation only in the optical and/or infrared bands that are likely to dominate the outer disc.  A valuable diagnostic would be searching for eclipses in H$\alpha$ or other emission lines, which are always produced in the outer accretion disc, but only produced from jets in rare sources like SS~433.

To estimate the size of this effect, we must know the radius of the outer accretion disk.  The outer accretion disk can vary in size over time, growing when during outbursts, and shrinking during quiescent periods \citep{Smak1984,Hameury2019}.   The circularization radius will represent a good lower limit to the size of the disc, and under some circumstances will be very close to the actual size of the outer disc.  The circularization radius will generically be the radius at which the angular momentum per unit mass of the accreted gas is equal to the angular momentum per unit mass of a circular orbit around the accretor.  The functional form will differ for wind-fed versus Roche lobe overflow systems.  We take the expressions for both from \citet{FKR2002}.

For wind-fed systems, the circularization radius will be:
\begin{equation}
\frac{r_{circ}}{a} = \frac{m_1^3(m_1 + m_2)}{16\lambda^4m_2^4}\left(\frac{r}{a}\right)^4,
\end{equation}
Here, the accretor is defined to have mass $m_1$ and the donor to have mass $m_2$ even though in some wind-fed systems, the donor will be the heavier object.  The ratio of the wind-speed to the escape speed from the donor star is given by $\lambda$, $a$ is the orbital separation and $r$ is the radius of the donor star.

For Roche lobe overflow systems, the circularization radius will be:
\begin{equation}
r_{circ}/a = (1+q)[0.5- 0.227~ {\rm log_{10}}~q]^4.
\end{equation}
The practical upper limit on this value is $(1-\frac{r}{a})$, and if the computed value is larger, then the disc should extend to the inner Lagrange point.  This will be the case for $q<0.008$, a value likely to be reached only for ultracompact X-ray binaries with black hole accretors.

For cases of winds slower than the orbital speeds, and donor stars which nearly fill the Roche lobes, the two expressions approach becoming similar.\footnote{This may be the case for period gap systems -- see e.g. \citealt{MP2013}}  For the typical mass ratios in low mass X-ray binaries ($q\sim0.1-0.5$), the circularization radius will be about 1/3 of the orbital separation.  The eclipse of jet by the outer disc should {\it not} show any orbital phase dependence unless the disk has a stable non-circular component.  This may be the case if there is a puffy rim at the hot spot where the accretion stream joins to the outer disc, but it is likely to be a weak effect. 

The maximum value for the outer radius is generally taken to be the {tidal radius of the accretion disc, which will be about 0.9 times the Roche lobe size of the} accretor \citep{FKR2002}.  If the outer disc extends that far out, then at mid-eclipse, the inner counterjet will be eclipsed all the way up to the height associated with the outer part of the donor star.  For the calculations we show here that illustrate the effects of discs eclipsing the inner counterjet, we will use $r_{circ}$ for the outer disc radius, but we caution that there is abundant evidence that some systems, some of the time, have outer discs that extend considerably further out and that the outer radii can be variable \citep{Whitehurst1988, Osaki2013}.

Assuming that the disc is circular, and defining its outer radius to be $r_{out}$, we find that the height at which the disc eclipses the {counterjet}, $h_{dej}$, is given by:
\begin{equation}
    h_{dej} < r_{out}~{\rm cot}~i. 
\end{equation}
  This assumes a fully planar disc, although the deviations from planarity of the disc matter are significant only for nearly edge-on systems.  

This facet of the data then gives an opportunity to estimate the height of the base of the jet, since the wavelength up to which which the disc eclipses the jet will scale linearly with $h_{dej}$, and will correspond to the frequency $\nu_{dej}$.  It will then be the case that:
\begin{equation}
    h_0 = \frac{r_{out}{\cot}~i~\nu_{dej}}{\nu_{br}},
\end{equation}
where $h_0$ is the height of the base of the jet.  We can then follow through with the requirement that $r_{out}\geq{r_{circ}}$.  From this, then $\tan~i~\nu_{dej} \nu_{br} = \frac{h_0}{r_{out}}$.  Then, since $r_{circ}$ can be estimated by determining the mass ratio via the eclipse duration at the dominant frequency for the eclipse, and the inclination angle can be determined from the range of frequencies eclipsed, the only unknowns are $h_0$ and the ratio of $r_{out}$ to $r_{circ}$. Therefore, $h_0$ can be constrained strongly from the eclipses as well.

For aligned jets, if the eclipses from both the outer disk and from the donor star are seen, then an extra constraint on the combination of the inclination angle and jet structure can be obtained.  Care must be taken, of course, to determine where the actual outer disc radius is located.  In many systems, the disc will spread outward from the circularization radius, and can be significantly larger than this value.  

The eclipsing by the outer disk will be present in spectroscopy of the jet alone.  Because other components of the accretion flow are very likely to emit in the same bands as the jet base, this may develop into a new approach for finding the jets' break frequencies more robustly than they can be found by spectral measures.  It will be especially important if there is a substantial contribution from a hot inner flow that will only be eclipsed if the whole thermal accretion disc is eclipsed.

\subsection{Eclipses of the outer disc by the star}
The outer disc can be eclipsed without the inner disc (i.e. the origin of the X-ray emission).  X-ray eclipses (excluding eclipses of scattered X-rays) will take place for
\begin{equation}
    {\rm cos}~i < \frac{r}{a}
\end{equation}
while eclipses of the outer disc will take place for:
\begin{equation}
    {\rm cos}~i < \left(\frac{r}{a-r_{out}}\right).
\end{equation}

For small mass ratios, this effect can substantially affect the critical inclination angle for seeing some type of eclipse.  Furthermore, when accretion discs are strongly irradiated, as in the outbursts of soft X-ray transients, the outer disc may dominate the total optical emission due to its larger surface area relative to the hotter inner disc, and these will turn out to be the situations where the outer disc is largest, as well.  These grazing eclipses, will, in many cases have relatively weak signatures for short durations, but they should be nearly as strictly periodic as the orbit itself (with perhaps some low-level jitter due to variations in the structure of the outer disks), and hence it should be possible to identify them over time in Fourier transforms of long, well-sampled light curves, as might be obtained by Evryscope, which obtains high cadence, high duty cycle photometry of a large fraction of the sky, making it ideal for indentifying grazing eclipses \citep{Law2015}. We do not consider this effect in the remainder of this paper, but do note that it represents a potential tool for identifying sources within the range of inclination angles where only the outer disc is eclipsed.

\section{The approaching jet}

One of the challenges of searching for jet eclipses will be that in the case of aligned jets, it is the counterjet that will be eclipsed.  When the jet speed is large, this can lead to a situation where the radio flux is dominated almost entirely by the approaching jet's flux, regardless of any effects of eclipses.  The Doppler factor, $\delta$ for the jet will be given by:
\begin{equation}
    \delta = \frac{1}{\Gamma (1-\beta {\rm cos~}i)}.
\end{equation}
If we take, for example the parameters for Cygnus~X-1 from \citep{Tetarenko2019}, with $\Gamma$=3 and $i$=27 degrees, we find that the $\delta$ for the approaching jet will be 2.1, while $\delta$ for the counterjet will be 0.18.  For flat spectrum steady jets, the change in the flux due to Doppler boosting is $\delta^2.22$, so the approaching jet will be more than 100 times as bright as the counterjet.  It is for this reason that we consider the effects of jet eclipses only for neutron stars and quiescent black holes, where the jets are likely to be significantly slower.  For black hole systems that are quite edge-on, the eclipses may still be easily measurable. {For a slower jet speed of $\Gamma$=1.5, the $\delta$ for the approaching jet will be 2, while the receding jet $\delta$ will be 0.4, so that if the power scales with $\delta^{2.2}$, the counterjet will have about 3\% of the total flux density.}

\section{Practical candidate targets and predictions for them}

A few practical considerations exist for testing the model.  These will result in the best targets being bright low mass X-ray binaries with neutron star accretors.  The other class of strong candidate targets will be the brightest quiescent black hole X-ray binaries.

First, the most effective wavebands in which to search for jet eclipses will be those in which the emission is unambiguously from the jet.  Radio emission will be suitable for this in nearly all systems.  In many systems (i.e. those with shorter orbital periods and hence fainter donor stars and weaker thermal emission from the outer accretion disk)  infrared emission may also be suitable for these purposes \citep{Migliari2006,Migliari2010}.  In cases where the accretion disk may be emitting at the same wavelength, and especially where the accretion disk's flux may be modulated by either the orbital period, or a very similar superhump period, it may be difficult to distinguish a period due to a jet eclipse from other forms of periodic modulation without very long time series.  As a counterbalance, for making use of radio observations, either relatively large orbital separations or very low inclination angles may be necessary  -- relatively edge-on systems with short periods will have their eclipses take place at higher frequencies.  Additionally, the eclipse durations as a fraction of the orbital period will be larger for systems with larger orbital period because the eclipse duration will scale as ${\rm sin}^{-1} r/a$, which increases with mass ratio.  Fortunately, in most systems, the emission region that is eclipsed is likely to be in the radio through submillimeter part of the spectrum, and this is confirmed in the simulations.

Second, the utility of the eclipses of the jet may be limited in systems where the donor star shows strong winds.  In these systems, the free-free absorption in the stellar wind can produce a strong modulation of the jet flux coming from a wide range of heights \citep{Brocksopp2002}, including heights at which the photosphere of the donor star is not in front of the jet.  This jet {flux} modulation may be useful for making measurements of the orbital periods even for systems which are so heavily extincted that the donor star cannot be detected, and it may be useful for establishing the existence of misaligned jets. The modulation is unlikely to be useful for understanding the jet's structure or measuring inclination angles or mass ratios.  Third, because in the aligned jet case, the counterjet will be the eclipsed, rather than the approaching jet, the slower neutron star jets will show stronger eclipses due to smaller effects from Doppler boosting and deboosting.  Fourth, a large fraction of accreting neutron stars are persistent sources or high duty cycle transients, while only one accreting black hole in the Galaxy with a low mass donor is persistent (4U~1957+11 -- \citealt{Nowak2012}) and only one other one is a high duty cycle transient (GX~339--4 -- \citealt{Buxton2012}).  Observations for the purposes of looking for jet eclipses will already be strongly time-constrained because they need to occur around zero phase. Searching persistent sources make the scheduling easier by allowing scheduling at any zero phase epoch, and the high duty cycle transients at least allow some flexibility. Furthermore, in many cases persistent sources allow the orbital phase to be determined independently of the searches for the eclipses.  A substantial fraction of persistent X-ray binaries have unknown orbital periods.  Additionally, in the neutron star cases, some of these show alternative evidence for being ultracompact X-ray binaries, for which long integration times will sample many orbital periods.   None of these issues precludes the idea that jet eclipses might eventually be used to understand black hole X-ray binaries; it merely argues that the neutron star X-ray binaries are better choices for establishing that the methodology works.

\subsection{Z-sources}

The Z-sources are the brightest low-magnetic field accreting neutron stars, both in the X-ray band and the radio band, so named because they trace out a shape that resembles the letter Z in X-ray color-color diagrams \citep{HasingervanderKlis}.  Most of the known Z-sources have been persistent accretors since the era of {\it Uhuru} or even since the earliest sounding rocket searches for X-ray sources.  The Z-sources tend to show longer orbital periods than most other accreting neutron stars, as well.  Longer orbital periods present the advantage that the eclipsed emission comes from further up the jet, in some cases pushing the eclipses into the cm-wave bands with which it is easiest to make measurements.  They present the disadvantage that the orbits are slower, meaning both that it takes more time per orbital period to cover the eclipse while also being likely to have intrinsic source variability dominate over measurement noise.

Three persistent Z-sources have known orbital periods: {Sco~X-1} (18.9 hours -- \citealt{Steeghs2002}), Cyg~X-2 (9.8 days -- \citealt{Casares2010}), and GX~349+2 (22 hours -- \citealt{Wachter1997}).  Of these systems, only Sco~X-1 has a reliable geometric parallax distance \citep{Fomalont2001}, so it is difficult to estimate the luminosities beyond the standard assumption that they are roughly $L_{EDD}$.  Fortunately, the expected relationship between $\nu_{br}$ and $L_X$ is that $\nu_{br}{\propto}L_X^{1/3}$ for radiatively efficient sources, so uncertainties of even a factor of 3 in luminosity lead to only a factor of 1.4 in $\nu_{br}$, likely smaller than some of the other uncertainties in the problem.  The inclination angles represent a more serious challenge, given that the cotangent can vary by a factor of 3 between 30 and 60 degrees.  Because there is neither a spectroscopic confirmation of the period of GX 349+2 nor any estimate of its inclination angle, we do not consider it here.  In Table \ref{Z-sources}, we summarize the key parameters for the systems.  

Scaling from 4U~0614+091, assuming 4U~0614+091 has a break frequency of $2\times10^{13}$ Hz \citep{Migliari2010}\footnote{The break frequency of a second neutron star X-ray binary, 4U~1728--34, has been inferred from ALMA data, but was not directly detected, and is within a factor of a few of this value for a source that is likely to be at a somewhat higher X-ray luminosity \citep{DiazTrigo2017}.}, an $L_X$ of $10^{36}$ erg/sec, we expect a break frequency at about $1-1.4\times10^{14}$ Hz for the two Z-sources.  We use $10^{14}$ for the purposes of Table \ref{Z-sources}).  Then, and assuming that the $h_0$ is about $6\times10^8$ cm, we find that the expected frequencies for the eclipses will be at 157 GHz for Sco~X-1, and 96 GHz for Cyg~X-2.  This frequency range can be probed by several facilities, such as ALMA, NOEMA, JCMT and the Large Millimeter Telescope.  Given the brightness of these systems (especially Sco~X-1), it should be practical to search for the jet eclipses in these Z-sources despite the challenges that usually arise from working in the millimeter band.  We anticipate that even after dilution due to the approaching jet, that the eclipse depth at about 150 GHz should be about 10\%, as illustrated for spectra at different orbital phases in Figure \ref{fig:sco} and for light curves for different frequencies in Figure \ref{fig:sco_lc}.

\begin{figure}
    \centering
    \includegraphics[width=3.5 in]{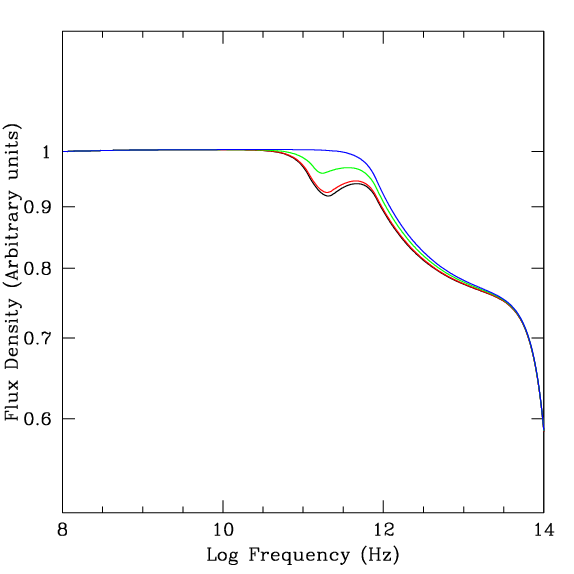}
    \caption{The jet spectra for Sco~X-1, taking both disc and stellar eclipses into account, and assuming $\beta=0.3$.  Each curve is at a different orbital phase.  From bottom to top, the phases are 0.01 (black), 0.02 (red), 0.04 (green), 0.07  (dark blue).  The dip just above $10^{11}$ Hz is due to the star eclipsing the jet, while the effect of the outer disc blocking the inner jet produces the curvature at $\approx10^{12.5}$ Hz.}
    \label{fig:sco}
\end{figure}

\begin{figure}
    \centering
    \includegraphics[width=3.5 in]{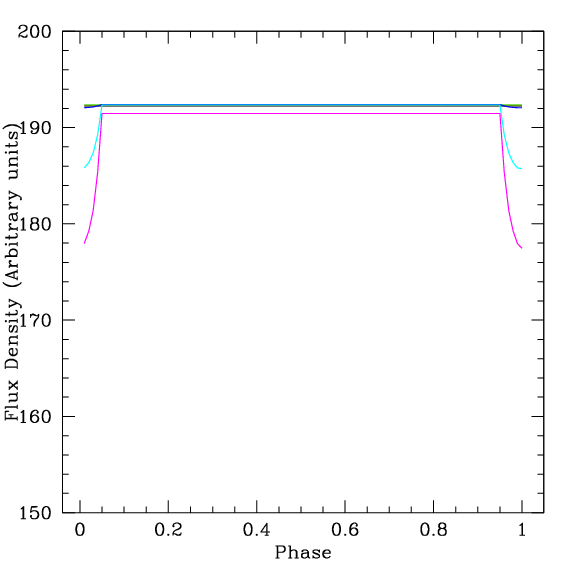}
    \caption{Model calculations for Sco~X-1's light curve.  From top to bottom, frequencies are: $10^9$ Hz (black), $3\times10^9$ Hz (red), $10^{10}$ Hz (green), $3\times10^{10}$ Hz (blue), $10^{11}$ Hz (cyan), and $3\times10^{11}$ Hz (purple).  The calculation here is for the full solution for the jet, including disc eclipses and dilution by the approaching jet.}
    \label{fig:sco_lc}
\end{figure}

Beyond this, the CMB-S4 \citep{CMBS4} project, if constructed, would have some powerful capabilities for studying bright source variability in the 30-300 GHz range.  How useful it would be for studying Sco~X-1 and the few other brightest radio sources in the sky for this depends on deployment decisions that have not yet been made.

If some of the systematic uncertainties in the problem lead to a lower range of frequencies at which the eclipses take place (which may be likely, given that the fraction of the power that goes into the jet is probably smaller for the slim discs at near Eddington luminosities than the {\it bona fide} thick discs at strongly sub-Eddington luminosities), then this would open up the possibility of using facilities at 40 GHz, like the Very Large Array and even the Very Long Baseline Array for brighter sources like Sco X-1.  With the Next Generation Very Large Array, it would be possible to use subarrays to cover the full range of frequencies from 1-100 GHz for sources in the mJy flux density regime.  A major caveat for the Z-sources is the assumption of the extrapolation of the atoll sources' break frequencies over a few orders of magnitude in luminosity.  A particularly important point is that in some epochs, the core of the radio emission from Sco~X-1 has been seen to be flat spectrum, while in others, it is seen to show a steep spectrum ($\alpha\approx{-0.6}$) -- in this case, these uncertainties are more likely to make the project easier to execute than harder, but they also increase the risk associated with any particular observing strategy.  

The strong and rapid variability of the Z-sources on a wide range of timescales will make this work challenging, nonetheless.  We believe that this problem can be dealt with by making the radio/mm observations simultaneously with X-ray observations.  The searches for eclipses can be made by stacking epochs near zero phase with similar X-ray properties.  This will necessitate collecting more data than might be required for sources that show steadier X-ray properties.  However, this may be offset by the fact that smaller arrays and even single dishes will be able to make these measurements.  

\begin{table}
\caption{Key data and predictions for the Z-sources.  The donor masses and inclination angles come from \citet{Steeghs2002} for Sco~X-1,  and \citet{Casares2010} for Cyg~X-2.}
\begin{tabular}{l|l|l|l|l|l}
Source     &  $P_{orb}$ & $a$ & i& $m_2$&$\nu_{ecl}$\\
\hline
&hours&cm&degrees&$M_{\odot}$ & GHz\\
\hline
Sco~X-1     &  18.9 & $3.0\times10^{11}$ & 38 & 0.42& 157 \\
Cyg~X-2& 236.3& $1.8\times10^{12}$ & 62.5 & 0.58& 96 \\
\hline
\end{tabular}
\label{Z-sources}
\end{table}

\subsection{Atoll sources}

For the atoll sources, we have a better handle on the input parameters.  The break frequencies are well characterized for 4U~0614+091, and 4U~0614+091 has also been established to show a flat spectrum longward of the break.  The radio/X-ray correlation for most of the atoll sources appears to show a $L_R \propto L_X^{1.4}$ relationship, meaning that close to the state transition between the island state and the banana state\footnote{The island state is the neutron star equivalent of the hard state for black holes, while the banana state is the equivalent of the soft state \citep{vanderklisbook}.}, the sources will be most powerful in the radio band.  These correlations show considerable scatter, but the major point of concern is just to identify the states where the radio emission will be brightest, and so it is not a major concern for our purposes whether the $L_R \propto L_X^{1.4}$ holds up with bigger samples with better measurements.  There is some additional evidence that some, but not all, of the atoll sources continue to be relatively bright radio sources even in the banana states (compare e.g. \citealt{Migliari2004} with \citealt{DiazTrigo2018}). 

We thus anticipate that the break frequencies, when the sources are bright enough to be detected, will always be within a factor of about 2 of $2\times10^{13}$ Hz -- this is the case for 4U~0614+091 \citep{Migliari2010} and for 4U~1728--34 \citep{DiazTrigo2017}.  For Aql~X-1, the break frequency in one observation in the hard intermediate state appears to be at about 100 GHz in some cases \citep{DiazTrigo2018}, but this is at 2$\sigma$, with only one data point above the break, and with data in the two bands which are not strictly simultaneous; at the present time, we regard the data from the two other sources as more representative of what is likely to be true with a larger sample of observations.  If we then take the height of the break to be $6\times10^8$ cm, we expect that the deepest eclipses will typically be at less than 30 GHz  when $a$~cot~i is greater than $4\times10^{11}$ cm.  For sources with inclination angles of 45$^\circ$, this will correspond to orbital periods of about 20 hours.  Alternatively, we could observe at higher frequency, or we could observe sources with lower inclination angles.

\subsubsection{Long period atolls}
The atoll sources which are at periods longer than 10 hours are Aql~X-1 \citep{Simon2002,Ootes2018}, 4U~1608--52 \citep{2004NuPhS.132..652S} and Cen~X-4.  The former two objects are transients which have frequent outbursts, while the latter source is a transient which had outbursts in 1969 and 1979 \citep{Evans1970,Kaluzienski1980}, but has not outbursted since then. For Aql~X-1, the system's mass ratio and radial velocity amplitude have been estimated from the donor's absorption lines, consistent with a range of $36^\circ - 47^\circ$ under the assumption that the neutron star must be between 1.2 and 3 $M_\odot$ \citep{MataSanchez2017}.  Taking the standard neutron star mass of $1.4M_\odot$, we find that $a=3.1\times10^{11}$ cm for Aql~X-1.  For a $40^\circ$ inclination angle, this would give a height up the jet of $3.7\times10^{11}$ cm for the longest duration of the eclipse, and hence the frequency associated with the maximum duration of the eclipse should be 33 GHz, a frequency already accessible with the VLA, and for which the ngVLA would make a large improvement in sensitivity.  When in X-ray bright hard states, Aql~X-1 tends to show radio emission at the $200-300 \mu$Jy level \citep{Tudose2009, Gusinskaia2020}.  The eclipse duration at the middle frequency should be 1.87 hours for the $q=0.41$ mass ratio of this system.  With the present VLA, in 1.87 hours, with 8 GHz of bandwidth, a noise level of 3 $\mu$Jy can be achieved at 30 GHz over an 8 GHz bandwidth.  In a narrower bandpass that might be used to search for the mid-eclipse eclipse, say 1~GHz (i.e. 3\% fractional frequency width), a noise level of 10 $\mu$Jy (about 3-5\% of the total flux density) would be achieved over the eclipse duration.  A small amount of time will be lost to calibration for such a project, but since the timing of the eclipse is known, the calibration observations could be conducted primarily outside the eclipse.

For Aql~X-1, at about 30~GHz, the counterjet can be expected to show a variation of about 25\% between the eclipsed and uneclipsed phases, and the approaching jet will lead to a dilution to a modulation of a bit more than 10\%.  Calculations of the spectra at different phases are shown in Figure \ref{fig:aql}, and calculations of the light curves are shown in Figure \ref{fig:aql_lc}.  It can also be seen that the deepest eclipse moves to lower frequencies as the eclipse progresses.  This is because of the asymmetry in the jet spectrum, in that the range of heights which contribute significantly to a particular wavelength is larger going upwards (where it falls off due to the $\nu^{-0.7}$ spectral shape) than downwards (where the self-absorbed spectrum goes as $\nu^{2.5}$).  

\begin{figure}
    \centering
    \includegraphics[width=3.5 in]{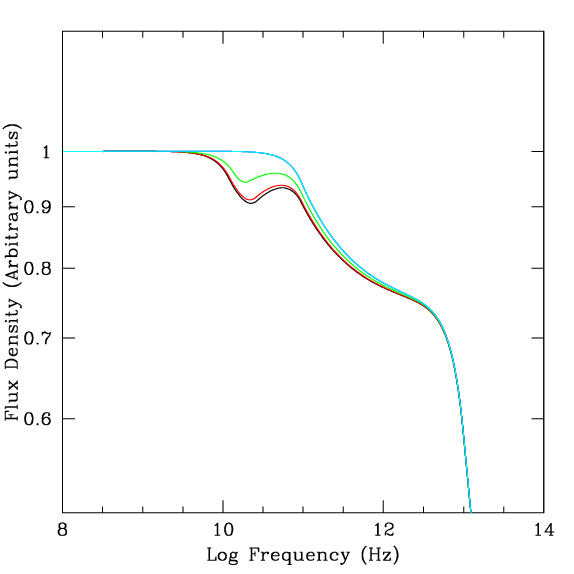}
    \caption{Model calculations for Aql~X-1.  From bottom to top, the orbital phases are: 0.01 (black), 0.02(red), 0.04 (green), 0.07(blue), and 0.14 (cyan).  The blue and cyan curves nearly overlap.  The calculation shows all relevant proceses, with both the jet and counterjet's contribution included, as well as occulation by both the disc and the donor star.}
    \label{fig:aql}
\end{figure}

\begin{figure}
    \centering
    \includegraphics[width=3.5 in]{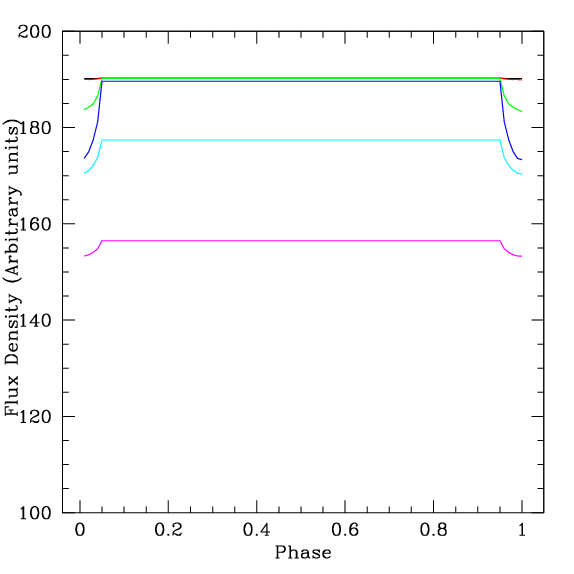}
    \caption{Model calculations for Aql~X-1's light curve.  From top to bottom, frequencies are: $10^9$ Hz (black), $3\times10^9$ Hz (red), $10^{10}$ Hz (green), $3\times10^{10}$ Hz (blue), $10^{11}$ Hz (cyan), and $3\times10^{11}$ Hz (purple).  The calculation here is only for the full solution for the jet, including disc eclipses and dilution by the approaching jet.}
    \label{fig:aql_lc}
\end{figure}

This object is a transient, but fortunately, it has a high duty cycle in outburst.  Aql X-1 typically has outbursts a few times a year, with bright ones about once a year \citep{Simon2002,Ootes2018}.  It may be that the brighter outbursts are superoutbursts \citep{MaccaroneReview}, an idea bolstered by the tentative discovery of superhumps from this system.  If so, the neutron star mass would likely be a bit higher than 1.4$M_\odot$, as the mass ratio for the system would need to be a bit smaller (i.e. $<$0.35).  Even the brightest phases of the "normal" \footnote{The normal outbursts are sometimes called mini-outbursts in the context of this source due to the fact that they were often missed before the system began to be monitored intensively.} outbursts should be bright enough to make these measurements.  Thus, there should be a few opportunities per year to detect the eclipses from Aql~X-1, and the opportunity to measure several eclipses per bright phase.  

This is not quite as easy a system to work with as the persistent systems, but it is, nonetheless, still achievable even with present technology.  With the ngVLA, such observations would be much easier, and subarraying could be done to cover a wider bandpass and help ensure that (1) the eclipsed frequency is actually covered and (2) that the putative eclipsed frequency range can be confirmed to be more strongly variable than adjacent frequencies.  This could allow higher quality measurements to be done without making as many repeated visits of the eclipsed phases in order to ensure that the phenomenon is periodic.

4U~1608-52 would be likely to show similar phenomenology -- it has a similar orbital period and is similarly a high duty cycle transient\citep{2004NuPhS.132..652S}.  Unfortunately, it is located too far south for VLA observations.  With the notable exception of ALMA, the Southern hemisphere radio facilities are not well suited for high frequency work, and this will not change with the first phase of the Square Kilometer Array.  This eclipse would likely fall within ALMA Band 1 once the Band 1 system is deployed \citep{DiFrancesco} and hence could potentially be done by ALMA, even in relatively poor weather.

\subsubsection{Short period atolls}

The shorter period atoll sources are likely to be eclipsed only at higher frequencies, unless they are viewed relatively face-on (in which case, they are most likely to be among the neutron star X-ray binaries with unknown orbital periods), or show strong superhump modulations (which can yield the orbital period to within a few percent, but cannot yield the orbital phase information).  A few of these systems are strong candidates to be viewed face-on -- e.g. Ser~X-1, 4U~1820--30 and GX~349+2 \citep{Cackett2008}.  The most interesting one is Ser~X-1.

Ser~X-1 shows two cycles of apparently periodic modulations of the motion of its emission lines, indicating a period of about 2 hours \citep{Cornelisse2013}.  It shows relatively narrow optical emission lines and a relatively small amplitude for the apparently periodic variations \citep{Cornelisse2013}, suggesting that the inclination angle is about 10 degrees.  It also shows a reflection spectrum that is indicative of a relatively face-on disc \citep{Matranga2017,Mondal2020}.  While ideally there would be a larger number of cycles of the orbital period covered, the sum of the evidence is suggestive of the idea that this is a face-on, short period binary.  If we take the parameters of \citet{Cornelisse2013}, then we find that the central location for the eclipse should be at about $3.7\times10^{11}$~cm, and hence at a frequency close to 30 GHz.  This system was one of the first soft state neutron star sources detected \citep{Migliari2004}, at a flux density of about 100 $\mu$Jy, and hence it should be a practical persistent source target if the orbital period and inclination angle are as assumed.  Given an expected mass ratio of 0.14 (assuming a 0.2 $M_\odot$ donor star and a 1.4 $M_\odot$ accretor), the source should be eclipsed about 7\% of the time at the central wavelength for the eclipse.

The large inclination angle allows the eclipse of a fairly large range of heights emitting at frequencies around 15 GHz.  The maximum flux loss during the eclipse is about 40\% of the flux from the counterjet. The flipside of this low inclination angle is that even for moderately relativistic jet speeds, the beaming may be significant, and the approaching jet may outshine the counterjet significantly. 
where $\delta$ is the Doppler boost factor.

For this system, $\delta_{app}$, the Doppler factor for the approaching jet will be 1.35, while the counterjet will have a Doppler factor of 0.77.  The approaching jet will be boosted to be about {3.4} times brighter than the counterjet before accounting for eclipses.  The strong eclipse of the counterjet will then correspond to a roughly {9\%} reduction in the total jet flux from the system, still fairly easily detectable with good signal-to-noise.  We plot the source spectra in figure \ref{fig:serspec}, taking into account the effects of both the approaching jet and the disc occulation of the counterjet, and the light curves at different frequencies in Figure \ref{fig:ser_lc}.

To detect such an effect from a 100 $\mu$Jy source with good statistical significance would require roughly 2$\mu$Jy sensitivity.  Reaching this flux density level at 15 or 30 GHz with the VLA requires about 4 hours of total exposure time.  A total of about 10 hours would thus be needed to do the project, to allow for similar exposure on and off eclipse and enough time for calibration.  At the present time, though, with the orbital period not yet confirmed, and the ephemeris completely unknown, the project would be considerably less efficient, as full orbits would be needed to allow the search for the period.  Still, with the program doable in 2 hour blocks, it could be gradually scheduled over time.  With the equatorial nature of the source, it would also make an excellent filler program for ALMA for times with relatively high water vapor where Band 1 observations are preferred.

\begin{figure}
    \centering
    \includegraphics[width=3.5 in]{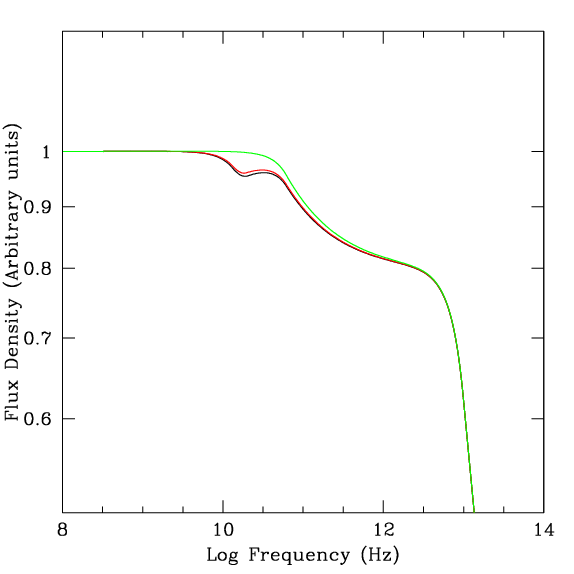}
    \caption{Model calculations for Ser~X-1.  From bottom to top, the orbital phases are: 0.01 (black), 0.02(red), 0.04 (green).  The calculation here includes both the approach and counterjet contributions, as well as the disc eclipse effects, assuming a jet speed of $0.3c$.  The first break, from flat spectrum to $\nu^{-0.05}$ is due to the eclipse of the inner jet by the outer disk, and the second break is where the innermost part of the approaching jet becomes optically thin.  Due to the low inclination used in the calculation, the disc eclipse of the jet is weaker, but spans a broader range of wavelengths than for most other sources.}
    \label{fig:serspec}
\end{figure}

\begin{figure}
    \centering
    \includegraphics[width=3.5 in]{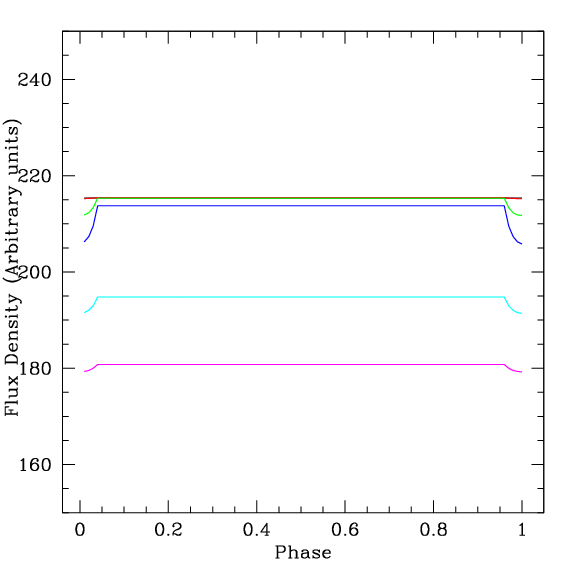}
    \caption{Model calculations for Ser~X-1's light curve.  From top to bottom, frequencies are: $10^9$ Hz (black), $3\times10^9$ Hz (red), $10^{10}$ Hz (green), $3\times10^{10}$ Hz (blue), $10^{11}$ Hz (cyan), and $3\times10^{11}$ Hz (purple).  The calculation here is only for the full solution for the jet, including disc eclipses and dilution by the approaching jet.}
    \label{fig:ser_lc}
\end{figure}

\subsection{Black hole systems}

\subsubsection{Long period systems: potential radio eclipsers}
While we focus on neutron star systems in this paper, if the jet speeds in the faintest black hole systems (especially quiescent systems) are sufficiently slow, the beaming effects may be relatively minor and the eclipses may be detectable.  For bright hard states, \citet{Casella2010,Tetarenko2019}, have found that the bulk Lorentz factor of the jet is likely to be $\gtrsim$3.  If the Lorentz factor is, in fact, as low as 3, then the counterjet eclipses should be detectable for many systems.  At $\Gamma=3$, the counterjet for a flat spectrum, steady jet should be only a factor of {about 4.5}  fainter than the approaching jet if the inclination angle is 70$^\circ$, roughly the threshold for eclipsing behavior for mass ratios typical of X-ray binaries.  At 28$^\circ$, however, $\delta_{app}$ will be 2, and the counterjet will have a $\delta=2/7$, so that the approaching jet will be {about 70} times brighter than the counterjet.  {In short period black hole X-ray binaries, when they are bright, we expect that the radio emission will typically come from well outside the orbital separation.  In longer period systems, such as Cygnus~X-1 \citep{Zdziarski2012Cygx1}, there does appear to be significant radio emission coming from within the orbit, illustrated by the strong modulation by the stellar wind, which is quite low density for spatial scales outside the orbital separation.}


{Given the likelihood that the quiescent binaries have slower jets (probably closer to $\beta=0.5$), they likely represent better targets}.  Most quiescent black hole X-ray binaries are too faint for the VLA to do this work, but the ngVLA (or, in the case of low frequencies at which the eclipse happens, the Square Kilometer Array) could potentially do these observations.  Additionally, the low mass ratios for black holes mean that the size of the eclipsed region will tend to be smaller than for the neutron star case.  Still a few sources merit some consideration as candidates for searches for jet eclipses.

V404~Cyg has orbital parameters of $P=6.5$~days, $m_1=12^{+3}_{-2} M_\odot$, $m_2=0.7^{+0.3}_{-0.2} M_\odot$, $i=56\pm4^\circ$, yielding an orbital separation of 34 $R_\odot$ and a donor radius of $6_\odot$, located at a distance of 2.4~kpc \citep{MillerJones2009}.  Its quiescent X-ray luminosity is about 3.9$\times10^{32}$ erg/sec and its quiescent radio flux density is typically about 200$\mu$Jy \citep{Corbel2008}, with the values corrected for the updated distance to the source.  We assume a break frequency of $4\times10^{12}$ Hz, based on a typical break frequency of $10^{14}$ Hz for a stellar mass black hole at $10^{37}$ erg/sec, and scaling by $L_X^{1/3}$, following \citet{HeinzSunyaev2003}.  We assume $h_0$ of $3\times10^9$cm.  Running our model with these parameters, we find that eclipses happen with a depth of 40\% with a central frequency of just over 10~GHz, which are diluted by an approaching jet with $\beta=0.5$ to have an eclipse depth of a bit less than 10\%.  The runs are in figure \ref{v404}, with light curves shown in figure \ref{fig:v404_lc}.  If $\Gamma=3$, then the approaching jet will have $\delta=0.72$ and the receding jet will have $\delta=0.22$, so the flux ratio should be {13.5}.  This would drop the 40\% amplitude eclipse of the counterjet down to about a {3\%} amplitude eclipse in total.  

 Another promising object is GRO~J1655--40, for which the system parameters are $P=2.6$ days, $i=68^\circ$, $m_1=5.4 M_\odot$, and $m_2=1.4 M_\odot$ \citep{Beer2002}; {notably, the jet inclination angle for this system is estimated to be 85$^\circ$ \citep{Hjellming1995}, and so this system is also a good test case for jet mis-alignment, and may furthermore have nearly half its flux contribution from the counterjet}.  These yield an orbital separation of $1.06\times10^{12}$ cm and a donor radius of $2.8\times10^{11}$ cm.  The X-ray luminosity for this source shows strong variations, with quiescent detections having been made at $2\times10^{31}-2\times10^{32}$ erg/sec \citep{Garcia2001}. Due to the larger mass ratio, the amplitude and duration of the eclipses for this system will be larger (with a factor of 3 separation), but given the more edge-on orbit and the closer separation, they will be at higher frequency (with the central eclipse at about 30~GHz).  The system is quite radio faint in quiescence, with an upper limit of 26 $\mu$Jy in the deepest observation of it that has been made \citep{Calvelo2010}.  It is thus unlikely to be a practical target for the VLA, unless, perhaps the existing upper limit was at an anomalously low flux level (which, given the factor of 10 variability in X-ray luminosity in quiescence, is a genuinely viable possibility), but it may be a good target for the ngVLA.  Its declination allows only very short observations from Northern Hemisphere facilities, which represents an additional challenge for both the VLA and ngVLA.  It may also potentially be a good target for ALMA Band 1 observations once they are commissioned.

\subsubsection{Short period systems: optical/infrared eclipsers}
The shorter period quiescent systems are likely to eclipse at much shorter wavelengths.  These may still be observable, especially in edge-on systems where the eclipses may then take place in the mid-infrared, and be in the JWST (or, for bright sources, SOFIA) window of observations.  Furthermore, if the sources are edge-on, they can be expected to have weaker effects from beaming. The most extreme source in this respect that is well characterized is XTE~J1118+480, although the short period dipping sources MAXI~J1659--152 \citep{Kuulkers2013} and Swift~J1357.2--0933 \citep{CorralSantana2013} may be even more extreme.  XTE~J1118+480 is both among the closest stellar mass black hole systems, with an orbital period of 4.1 hours \citep{GonzalezHernandez14}, and is also nearly edge-on, with an inclination angle of 68$^\circ$-79$^\circ$ \citep{Khargharia2013}.  As a result, the "bottom" of the donor star should come within a few times the height of the base of the jet.  The system is quite faint in quiescence, with $L_X = 3.5\times10^{30}$ erg/sec \citep{McClintock2003}. Here, there are some debates about whether the jet may extend as a flat spectrum source well into the ultraviolet in outburst, despite the outburst being considerably fainter than for most X-ray binaries.  The eclipse will start to become detectable at $\sim1/3$ of the break frequency, so the assumption about the break frequency is quite important.  If we take the standard scaling from the other sources, then we expect that the eclipses will be in the Terahertz regime in quiescence (and hence likely undetectable for the foreseeable future due to poor sensitivity in that band).  In outburst, though, the eclipses would be in the mid-infrared band.  If we instead take the viewpoint that the jet extends well into the ultraviolet in outburst, then we would expect to see jet eclipses in the optical band.  Given that the low mass ratio makes the phase range for the eclipse small, it is unlikely that sufficient data have been collected in the right bands during past outbursts, but it would be valuable to consider that possibility, regardless.

\begin{figure}
    \centering
    \includegraphics[width=3.5 in]{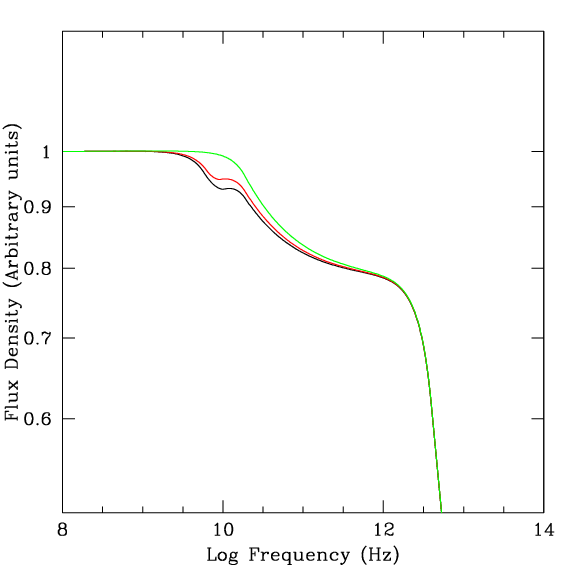}
    \caption{Model calculations for V404~Cyg.  From bottom to top, the orbital phases are: 0.01 (black), 0.02(red), 0.04 (green).  The spectrum plotting includes both the eclipse by the star and the occulation by the outer disc, and assumes an approaching jet with a speed of $0.5c$.  The eclipse by the star takes place at about $10^{10}$ Hz, while the occulation by the outer disc takes place around $10^{11}$ Hz.}
    \label{v404}
\end{figure}

\begin{figure}
    \centering
    \includegraphics[width=3.5 in]{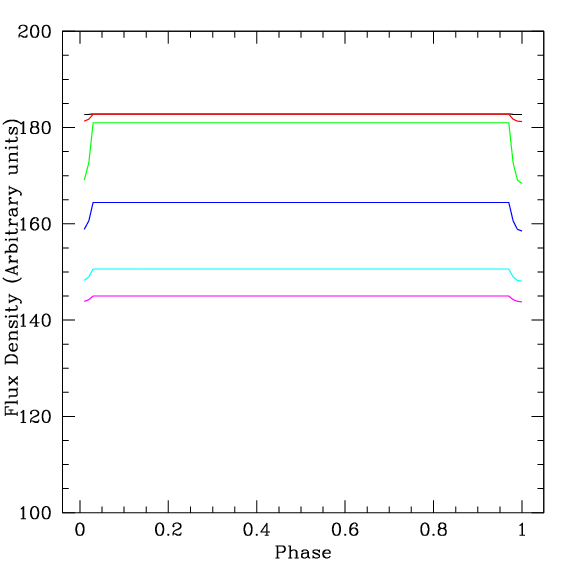}
    \caption{Model calculations for V404~Cyg's light curve.  From top to bottom, frequencies are: $10^9$ Hz (black), $3\times10^9$ Hz (red), $10^{10}$ Hz (green), $3\times10^{10}$ Hz (blue), $10^{11}$ Hz (cyan), and $3\times10^{11}$ Hz (purple).  The calculation here is only for the full solution for the jet, including disc eclipses and dilution by the approaching jet, under the assumption of a jet speed of $0.5c$ in quiescence.}
    \label{fig:v404_lc}
\end{figure}

\subsection{Cataclysmic variable jets: likely not to eclipse in the optically thick regime}

Over the past decade, it has become increasingly clear that cataclysmic variables show evidence for synchrotron emitting jets \citep[e.g.,][]{Kording2008,Russell16,Coppejans2020}.  Radio emission from these sources has been associated with both the compact jets and the transient jets. Radio flaring during the outbursts from these sources has been associated with the transient jets, with rapidly varying spectral indices indicative of emission evolving from optically thick to optically thin as plasmons expand \citep{Mooley2017,Fender2019}. Such an expanding plasmon was possibly resolved far outside the binary orbit \citep{Russell16}, however, the detection was only treated as tentative. Steadier emission from the compact jet at other times in the outbursts can show minute-timescale variability \citep[e.g.,][]{Russell16,Mooley2017,Fender2019}, while the spectral indices of these states generally have large uncertainties due to their radio faintness \citep{Coppejans2016}.  The "nova-like", persistently bright cataclysmic variables often show strong circular polarization, suggesting that at least some of these systems have at least a substantial part of the radio emission originating from a coherent process \citep{Coppejans2015}.  The symbiotic stars, which often show much brighter, steadier radio emission and have longer orbital periods (see e.g. \citealt{Seaquist1990}).  For them, the free-free absorption in the stellar wind would likely lead to "fuzzy" eclipses, if any eclipses are seen.

The minute-timescale variability for the non-flaring observations of both the dwarf novae and the novalike cataclysmic variables, combined with expected outflow speeds similar to the white dwarf escape speed (bolstered in part by the energetics arguments in \citealt{Fender2019} for the flaring sources), indicates that a significant fraction of the radio emission is likely to come from spatial scales of order $10^{11}$ cm, comparable to the orbital size scale.  The empirical data on these sources' characteristic size scales and jet spectra does not presently merit detailed predictions, particularly considering the faintness of the sources during their non-flaring states. However, searches for eclipses from these sources may prove fruitful, especially given that for many of them the orbital parameters are well-measured.

\section{Approaches to testing different jet models and future considerations}

In the discussion above, we have considered a particular form of the jet model -- one where the base is located at $3\times10^9$ cm and where the spectrum is flat.  Some systems clearly show non-flat spectra.  Additionally, some theoretical work takes a much smaller characteristic height for the jet base (e.g. \citealt{Markoff2005}), at least for some black hole systems.  Given that in the best cases (e.g. ultracompact X-ray binaries -- \citealt{Migliari2006}) the jet break can be estimated directly from the spectral energy distribution, for these objects, changing the height of the jet base by a factor of $10-100$ alters the frequency where the jet is eclipsed by the same factor.  Detection of the jet eclipses can thus potentially test these models for the location of the height of the break very effectively.  We do not present plots for models with lower values of $h_0$, as they will change primarily by a shift in the wavelengths at which phenomena occur, and not in the qualitative appearance of the eclipses.

We can also test models for inverted spectra using eclipses.  The simplest means for producing an inverted spectrum from a jet is to have the power in random kinetic energy drop slightly moving up the jet.  This will, in turn, also lead to deviations from linearity in the synchrotron self-absorption frequencies.  In principle, this could also be produced by changes in the Doppler beaming as one moves up the jet.  Most X-ray binaries are viewed at angles such that both the approaching and counterjets are deboosted -- for a viewing angle of 70$^\circ$, the approaching jet has $\delta<1.06$ for all speeds and $\delta<1$ for speeds above about 0.6$c$), while for a 45$^\circ$ inclination angle, $\delta<1.4$ for all angles, and $\delta<1$ for $\beta>0.94$, which corresponds to $\Gamma=2.8$.  Thus, for typical cases, for the black holes to have inverted spectra due to acceleration, it would require that they have jets which are accelerated, so that the Doppler deboosting is less important in the inner regions than in the outer regions.  Because a broad range of parameters space is possible here, it is likely that the jet eclipses in conjunction with timing of the jets  \citep{Casella2010,Gandhi2017,Tetarenko2019} will provide the solution to the problem.

At the present time, the model presented here makes predictions only for the case of aligned jets.  For the case of neutron stars in low mass X-ray binaries, where the spin of the neutron star is produced through accretion torques, this is unlikely to be a large problem.  For black hole jets, it may be a serious problem.  The jet is also assumed to be line-like in this model.  Eventually, we will wish to consider its transverse structure, but it is likely that it will be possible to study these only with extremely high signal-to-noise.

\subsection{Sub-arraying and the Next Generation Very Large Array}

Moving forward, especially when working with fainter sources, the Next Generation Very Large Array (ngVLA) \citep{ngvla} will be essential for allowing this technique to be applied to a large number of sources.  Its frequency range matches up well with where a large fraction of sources show eclipses.  Furthermore, its wide bandwidth for each receiver, along with its large number of dishes, allowing sub-arrays to be deployed to cover the full 1-100 GHz radio spectrum all at once with good sensitivity, will mitigate the problematic effects of source variability, and the issue of having the frequencies for the deepest eclipses being unknown before the data are collected.  Sub-arraying with the Karl G. Jansky Very Large Array has already been demonstrated to open up new capabilities in measuring jet variability \citep{Tetarenko2017}, and the ngVLA will be far more powerful in this respect.  The Square Kilometer Array, the other major new planned radio facility, lacks the high frequency capability to conduct this type of work for all but the longest period or lowest break-frequency systems.






\section{Conclusions}

Because the jets from X-ray binaries extend well outside the binary orbits, the jets will always be eclipsed over some range of orbital phase, even when the accretion disc is not eclipsed.  These eclipses present a powerful tool for probing both the nature of the jet, and understanding the binary system parameters.  In this paper, we have shown that, for many systems, these eclipses should be detectable with observations with current day facilities, if carefully designed observational campaigns are made.

\section*{Data Availability Statement}

No new observational data are presented in this paper.  The code used for this work is available at \url{https://github.com/tjmaccarone/jeteclipse}.

\section*{Acknowledgements}

TJM thanks the Anton Pannekoek Institute for hospitality while some of the foundational discussions for this project were conducted.  We also thank Greg Sivakoff, Chris Britt, Mar\'ia D\'iaz Trigo, and James Miller-Jones for useful discussions. JvdE and ND are supported by a Vidi grant from the Netherlands Organization for Scientific Research (NWO), awarded to ND.  {We thank the anonymous referee for rapid reports, in challenging times, that improved the quality of the paper.}





\bibliographystyle{mnras}
\bibliography{mnras_template} 







\bsp	
\label{lastpage}
\end{document}